\documentclass[12pt]{article}

\ifx\pdfoutput\undefined
\usepackage[dvips,bookmarks]{hyperref}
\else
\usepackage{hyperref}
\fi
\hypersetup{colorlinks=false,bookmarksopen,bookmarksnumbered,citecolor=blue,
   pdfstartview=FitH}

\usepackage[dvips]{graphicx}
\usepackage{latexsym}
\usepackage{amssymb,amsfonts,amsmath}
\usepackage{graphicx}
\usepackage{indentfirst}
 \usepackage{bbm}
\usepackage{mathrsfs}

\usepackage{amsmath, amsthm,amssymb}
\usepackage{mathrsfs}
\usepackage{hyperref}
\usepackage{amsfonts}
\usepackage{dsfont}
\usepackage{slashed, tensor}
\usepackage{booktabs}
\usepackage{graphicx}
\usepackage{mathrsfs}

\parskip=\medskipamount

\newcommand {\cL}{{\cal L}}

\newcommand {\cN}{{\cal N}}

\def\a{\alpha}

\def\b{\beta}

\def\d{\delta}

\def\g{\gamma}

\def\q{\theta}

\def\s{\sigma}

\def\O{\Omega}

\def\rd{{\rm d}}
\def\ri{{\rm i}}

\newcommand{\ve}{\varepsilon}                            

\newcommand{\pa}{\partial}                           
\newcommand{\hf}{\frac12}

\newcommand{\be}{\begin{equation}}
\newcommand{\ee}{\end{equation}}
\newcommand{\bea}{\begin{eqnarray}}
\newcommand{\eea}{\end{eqnarray}}
\newcommand{\non}{\nonumber}
\newcommand{\ba}{\begin{array}}
\newcommand{\ea}{\end{array}}

\newcommand{\1}{\underline{1}}
\newcommand{\2}{\underline{2}}

\newcommand{\ju}{\underline{j}}
\newcommand{\ku}{\underline{k}}

\newcommand{\bm}[1]{\mbox{\boldmath$#1$}}

\def\double #1{#1{\hbox{\kern-2pt $#1$}}}

\newcommand{\sSp}{\mathsf{Sp}}
\newcommand{\sSU}{\mathsf{SU}}
\newcommand{\sSL}{\mathsf{SL}}
\newcommand{\sGL}{\mathsf{GL}}
\newcommand{\sSO}{\mathsf{SO}}
\newcommand{\sUSp}{\mathsf{USp}}
\newcommand{\sO}{\mathsf{O}}
\newcommand{\sU}{\mathsf{U}}

\newcommand{\sOSp}{\mathsf{OSp}}

\newcommand{\bsubeq}{\begin{subequations}}
\newcommand{\esubeq}{\end{subequations}}

\numberwithin{equation}{section}

\begin{document}

\begin{titlepage}
\begin{flushright}
May, 2016
\end{flushright}
\vspace{2mm}

\begin{center}
{\Large \bf  Implications of $\bm{\cN=5, 6}$ superconformal symmetry\\[2mm]
in three spacetime dimensions
}\\
\end{center}

\begin{center}

{\bf
Sergei M. Kuzenko and Igor B. Samsonov\footnote{On leave from Tomsk Polytechnic University, 634050
Tomsk, Russia.}
}

{\footnotesize{
{\it School of Physics M013, The University of Western Australia\\
35 Stirling Highway, Crawley W.A. 6009, Australia}} ~\\
}

\end{center}

\begin{abstract}
\baselineskip=14pt
For general $\cN=5$ and $\cN=6$ superconformal field theories in three dimensions, 
we compute the three-point correlation functions of the supercurrent multiplets. 
In each case, $\cN=5$ and $\cN=6$, the functional form of this correlator 
is uniquely fixed modulo an overall coefficient which is related, by superconformal Ward identities, to the parameter in the two-point function of the supercurrent. 
The structure of the correlation functions obtained is consistent 
with the property  that every $\cN=5$ superconformal field theory, 
considered as a special $\cN=4$ theory, is invariant under the mirror map. 
\end{abstract}

\vspace{1cm}
\begin{flushright}
{\it Dedicated to the memory of Professor Nikolay G. Pletnev}\\[5mm]
\end{flushright}

\end{titlepage}

\newpage
\renewcommand{\thefootnote}{\arabic{footnote}}
\setcounter{footnote}{0}

\tableofcontents
\vspace{1cm}
\bigskip\hrule

\allowdisplaybreaks


\section{Introduction and summary}

This is a continuation of our recent works \cite{BKS1,BKS2} in which 
the two- and three-point correlation functions of the supercurrents and flavour current multiplets in three-dimensional (3D) $\cN$-extended
superconformal field theories have been 
computed for the cases $1\leq \cN \leq 4$. 
The present paper extends the results of  \cite{BKS1,BKS2} 
to the  3D field theories possessing $\cN=5$ \cite{Hosomichi2008} and $\cN=6$ 
\cite{ABJM,ABJ} superconformal symmetry. The $\cN=6$ superconformal field theories 
are often referred to as the ABJ(M) theories.\footnote{These theories possess
the  remarkably simple formulation \cite{BILPSZ}
in $\cN=3$  harmonic superspace \cite{ZH}. 
}

Although the family  of  $\cN$-extended superconformal field theories 
in three dimensions
is very large for $\cN\leq4$, 
it becomes much smaller for  $5\leq\cN\leq 8$. 
The latter families invariably consist of  superconformal
Chern-Simons theories interacting with supersymmetric matter in the bi-fundamental representation of the gauge group $G$
such that the amount of supersymmetry depends on the choice of $G$.
The allowed gauge groups are as follows:
 $\sSp(2M)\times \sO(N)$ for  $\cN=5$ 
  \cite{Hosomichi2008},  
  $\sU(M)\times \sU(N)$ or $\sSp(2M)\times \sO(2)$  $\cN=6$ for
  \cite{Hosomichi2008,ABJM},
   and  only the gauge group
    $\sSU(2)\times \sSU(2)$ for $\cN=8$
     \cite{BL1,BL2,Gus}. 
Clearly the range of $\cN=5$ and $\cN=6$ superconformal field theories 
are still pretty wide, and their properties are known to be quite fascinating.

In supersymmetric field theory in $d$ dimensions, 
the supercurrent \cite{FZ} is a supermultiplet containing
 the energy-momentum tensor and the supersymmetry currents, along with
 some additional components such as the $R$-symmetry current.
Thus the supercurrent contains fundamental information about the symmetries of
every supersymmetric field theory.
In the case of 3D extended superconformal field theories with $\cN>4$, 
 the supercurrent was introduced in \cite{BKNT,KNT} (see also \cite{BKS1}). 
 It is described by a primary real 
$\sSO(\cN)$ four-form
 superfield 
 $J^{IJKL}=J^{[IJKL]}$ of dimension 1, 
 $I=1, \dots, \cN$. 
The conformal supercurrent is  subject to the conservation equation 
\bea
D_{\a}^I J^{JKLP} = D_\a^{[I} J^{JKLP]}
- \frac{4}{\cN - 3} D_{\a}^Q J^{Q [JKL} \d^{P] I} ~ ,
\label{1}
\eea
where $D_\a^I$ denotes the spinor covariant derivative.
In the $\cN=5$ case, it is convenient to replace the four-form $J^{IJKL} $
with its Hodge-dual one-form $J^I$ defined by 
\bea 
J^I :=  \frac{1}{4!} \ve^{IJKLP} J_{JKLP} ~.
\eea
In terms of $J^I$, the conservation equation \eqref{1} turns into 
\bea
D^{(I}_\alpha J^{J)} - \frac15 \delta^{IJ} 
D^{Q}_\alpha J^Q = 0~.
\label{J-eq}
\eea
In the $\cN=6$ case, it is useful to switch from $J^{IJKL}$
to its Hodge-dual two-form $J^{IJ}$ 
\bea 
J^{IJ} := \frac{1}{4!} \ve^{IJKLPQ} J_{KLPQ}~.
\eea
In terms of $J^{IJ}$, the conservation equation \eqref{1} turns into 
\bea
D^I_\alpha J^{JK}= D^{[I}_\alpha J^{JK]}
-\frac25 D^Q_\alpha J^{Q[J}\delta^{K]I}~.
\label{N6-conserv}
\eea
A remarkable property of  $\cN=6$ supersymmetry 
in three dimensions is that  the supercurrent conservation equation
\eqref{N6-conserv} coincides with the Bianchi identity for an Abelian vector multiplet
\cite{KNT}. 

An important feature of the 3D 
extended superconformal theories with $\cN\geq5$ 
is the non-existence of conserved flavour current multiplets.
This point has recently been discussed in \cite{KKT}, 
and here we follow almost verbatim the discussion in \cite{KKT}.
In supersymmetric field theory in $d$ dimensions, 
  the conserved current multiplet is defined to be a supermultiplet 
containing a {\it single} conserved vector current $V^a$
(equivalently, a closed $(d-1)$-form), $ \pa_a V^a =0$,
along with some other scalar and spinor components. 
In three dimensions, one may think of a conserved current $V^a$
as the Hodge dual of the gauge-invariant field strength $F = \rd A$ 
of a gauge one-form $A$. 
For this reason an $\cN$-extended conserved current multiplet 
may be characterised by the same superfield type and 
the differential constraints as the field strength of an $\cN$-extended 
Abelian vector multiplet 
\cite{Siegel,HitchinKLR,ZP,ZH,Samtleben,KLT-M11}.\footnote{The
 conserved current multiplets with $\cN \leq 4 $ were reviewed in \cite{BKS1,BKS2}.} 
Thus for $\cN>2$, the conserved current multiplet should be defined 
 to be  a real antisymmetric superfield,
$L^{I J} = - L^{JI}$, constrained by 
\bea
D_{\a}^{I} L^{ J K}&=&
D_{\a}^{[I} L^{ J K]}
- \frac{2}{\cN-1}  D_{\a }^L L^{ L[J} \d^{K] I}
\label{3.4}
~.
\eea
For $\cN>4$, it turns out that the off-shell multiplet constrained by \eqref{3.4} 
possesses more than one conserved current at the component level.  
Moreover, it also  contains higher spin conserved 
currents for $\cN>4$ \cite{GGHN,KN}. Indeed,  for $\cN=6$
the conservation equation \eqref{3.4} 
coincides with  the supercurrent conservation equation
\eqref{N6-conserv}.  As a consequence, 
$L^{IJ} $ contains a symmetric, traceless and conserved  energy-momentum tensor 
$T^{ab}$
\bea
T^{ab}= T^{ba}~, \qquad \eta_{ab}T^{ab}=0~, \qquad \pa_b T^{ab}=0~,
\eea
In the $\cN=5$ case, $L^{IJ} $ contains a  conserved  symmetric spinor current
$S_{\a\b\g} =S_{(\a\b\g)}$ (supersymmetry current) defined by 
\bea
S_{\a\b\g} \propto \ve_{IJKLM} D^I_\a D^J_\b D^K_\g L^{LM}~.
\eea

In this paper we find the most general expressions for 
the two- and three-point correlation functions of the $\cN=5 $ and
$\cN=6$ supercurrent multiplets, which are allowed by the superconformal symmetry and are compatible with the conservation equations \eqref{J-eq} 
and \eqref{N6-conserv}, respectively. 
We show that the functional form of each of these correlators is determined by these requirements
modulo a single overall coefficient to be denoted by $c_\cN$ for the two-point functions and $d_\cN$ for the three-point ones. The ratio of these coefficients turns out to be fixed by the superconformal Ward identities. 

Every $\cN=5$ or $\cN=6$ superconformal field theory is a special $\cN=4$ superconformal field theory. 
It is of interest to understand what additional restrictions 
on the structure of $\cN=4$ correlation functions are implied by the  $\cN>4$ extended superconformal symmetry. 
For this we consider the reduction to $\cN=4$ superspace of the obtained 
correlators of the $\cN=5$ and $\cN=6$ supercurrents and compare them with the results of the work \cite{BKS2}. It is worth recalling that in general 
$\cN=4$ superconformal field theories the supercurrent three-point function  
has two linearly independent functional structures with 
free coefficients $d_{\cN=4}$ and $\tilde d_{\cN=4}$. We demonstrate that one of these coefficients is equal to zero  for all $\cN$-extended superconformal field theories 
with $\cN>4$, $\tilde d_{\cN=4}=0$.
In general,  it was shown in  \cite{BKS2} that 
$\tilde d_{\cN=4}$ is non-zero in those $\cN=4$ superconformal theories
which are not invariant under the mirror map. 
As discussed in \cite{BKS2}, $\tilde d_{\cN=4}$ is proportional to the difference of the numbers of left and right hypermultiplets \cite{BKS2}. 
However, it will be demonstrated in section \ref{sec3.1} that 
every  $\cN>4$ theory has an equal number of left and right hypermultiplets in the same representation of the gauge group. As a consequence, all 
$\cN>4$ superconformal field theories are invariant under the mirror map. 

As a by-product of the $\cN=4$ superspace reduction of the $\cN=5,6$ supercurrent correlation functions, we obtain new correlators in the $\cN=4$ superspace which correspond to conserved currents of extended supersymmetry and $R$-symmetry. These results are presented in the next two sections, which are devoted to the $\cN=5$ and $\cN=6$ theories, respectively.

In the study of correlation functions we follow the superspace approach which was 
originally elaborated for 4D $\cN=1$  superconformal field theories in \cite{Osborn} 
and generalised to the 4D $\cN=2$ case in \cite{KT}. 
For 3D superconformal field theories  this approach 
was originally developed in \cite{Park} and recently applied in \cite{BKS1,BKS2} to
 study correlation functions of supercurrents and flavour current multiplets. 
 In the present paper we use the superconformal formalism and definitions 
 introduced in  our works \cite{BKS1,BKS2}, which are somewhat different 
 from those  adopted in \cite{Park}. The summary of our definitions is given in the Appendix.
 

\section{$\cN=5$ superconformal field theories}

General $\cN=5$ superconformal field theories are 
supersymmetric Chern-Simons-matter models with appropriately 
chosen interaction potential \cite{Hosomichi2008,ABJ}. 
In this section we compute the two- and three-point correlation functions of supercurrents in such theories and consider their 
reduction to $\cN=4$ superspace.

\subsection{Correlators of $\cN=5$ supercurrent}

As discussed in section 1,  the $\cN=5$ supercurrent is  a primary 
dimension-1 superfield $J^I$ 
obeying the conservation equation \eqref{J-eq}.
The two-point function of the supercurrent,  which is compatible with this conservation law,  reads
\be
\label{N5-two-point}
\langle J^I(z_1) J^J(z_2) \rangle = c_{\cN=5}
\frac{u_{12}^{IJ}}{{\bm x}_{12}{}^2}~,
\ee
where $c_{\cN=5}$ is a free coefficient. Using the explicit form of the two-point structures ${\bm x}_{12}{}^2$ and $u_{12}^{IJ}$ given by (\ref{4.133}) and (\ref{two-point-u}), respectively, it is not hard to check that (\ref{N5-two-point}) obeys (\ref{J-eq})
at separate points, $z_1 \neq z_2$.

We look for the three-point function 
$\langle J^I(z_1) J^J(z_2) J^K(z_3) \rangle $
in the form
\be
\langle J^I(z_1) J^J(z_2) J^K(z_3) \rangle 
= \frac{u_{13}^{II'}u_{23}^{JJ'}}{{\bm x}_{13}{}^2 {\bm x}_{23}{}^2} H^{I'J'K}({\bm X}_3,\Theta_3)~,
\label{ans-N5}
\ee
where $H^{IJK}$ is a tensor depending on the three-point structures (\ref{three-points}). Since the supercurrent has dimension 1, this tensor should have the following scaling property
\be
H^{IJK}(\lambda^2 {\bm X},\lambda\Theta) = \lambda^{-2}
 H^{IJK}({\bm X},\Theta)~,
\label{scalingH}
\ee
for a real positive $\lambda$.

The supercurrent conservation law (\ref{J-eq}) implies that 
the tensor $H^{IJK}$ obeys the differential equation
\be
{\cal D}_\alpha^{(I} H^{J)KL}
-\frac15 \delta^{IJ}{\cal D}^M_\alpha H^{MKL} =0~,
\label{H-eq}
\ee
where ${\cal D}^I_\alpha$ is the generalised covariant derivative 
(\ref{generalized-DQ}).

The ansatz (\ref{ans-N5}) can be applied in different orders of operators $J^I(z_1)$, $J^J(z_2)$ and $J^K(z_3)$. In particular, interchanging the role of operators $J^I(z_1)$ and $J^J(z_2)$ in the correlator (\ref{ans-N5}) one finds the following symmetry property of $H^{IJK}$
\be
H^{IJK}({\bm X},\Theta) = H^{JIK}(-{\bm X}^{\rm T}, - \Theta)~.
\label{sym-1-2}
\ee
Similarly, swapping the operators $J^I(z_1)$ and $J^K(z_3)$ in (\ref{ans-N5}), one uncovers  the constraint
\be
H^{IJK}(-{\bm X}_1^{\rm T},-\Theta_1)
={\bm x}_{13}{}^2 {\bm X}_3{}^2
u_{13}^{JL}U_3^{LJ'} u_{13}^{II'}
u_{13}^{KK'} H^{K'J'I'}({\bm X}_3,\Theta_3)~.
\label{H-sym-1-3}
\ee
This equation was derived with the help of identities (\ref{222}).

We find the general solution of the equations (\ref{scalingH})--(\ref{sym-1-2}) 
in the form:
\be
H^{IJK} = 4 d_{\cN=5} \varepsilon^{IJKLM}\frac{A^{LM}}{X^3}
+d_{\cN=5} 
(\delta^{IJ}\varepsilon^{KLMNP} 
-\delta^{IK} \varepsilon^{JLMNP}
-\delta^{JK} \varepsilon^{ILMNP})
\frac{A^{LM}A^{NP}}{X^5}~,
\label{H-N5}
\ee
where $d_{\cN=5}$ is a free coefficient and the matrix $A^{IJ}$ is defined in (\ref{A}).
It is possible to show that the tensor (\ref{H-N5}) obeys also the equation (\ref{H-sym-1-3}) which ensures the invariance of the correlation function under the 
replacement $J^I(z_1) \longleftrightarrow J^K(z_3)$. In checking this, it is useful to express (\ref{H-N5}) in terms of the orthogonal matrix $U^{IJ}$ given in (\ref{U-explicit})
\be
H^{IJK}=-\frac1{12}d_{\cN=5}( H_1^{IJK}
-8 H_2^{IJK} + H_3^{IJK})~,
\label{tilde-H-N5}
\ee
where 
\begin{subequations}
\label{ids-Hi}
\bea
H_1^{IJK}&=&(\delta^{IK} 
 \varepsilon^{JPQRS}+\delta^{JK} \varepsilon^{IPQRS}+ U^{IJ} 
 \varepsilon^{KPQRS})\frac{U^{PQ}U^{RS}}{{\bm X}}~,
 \\
H_2^{IJK} &=&(U^{IL}\varepsilon^{JLKPQ}  
 - U^{LJ}\varepsilon^{ILKPQ}
 - \varepsilon^{IJKPQ} ) \frac{U^{PQ}}{\bm X}~, 
 \\
H_3^{IJK}&=& \varepsilon^{IJKLM}\frac{U^{LM}\Theta^4}{{\bm X}^3}~.
\eea
\end{subequations}
Here we have used the following identities:
\bea
&&\quad\qquad\qquad\qquad\qquad\qquad\qquad\ 
{\bm X}^2 = X^2 +\frac14 \Theta^4~,
\label{X2}
\\
&&(\varepsilon^{ILKMN}A^{LJ}+\varepsilon^{JLKMN}A^{LI})A^{MN}
=\frac14(\delta^{IK}\varepsilon^{JMNPQ}
+ \delta^{JK}\varepsilon^{IMNPQ}
\non\\&&\qquad\qquad\qquad\qquad\qquad\qquad\qquad\qquad
-2 \delta^{IJ}\varepsilon^{KMNPQ})
A^{MN} A^{PQ}~,\\&&
A^{IJ}\varepsilon^{KLMNP}A^{LM} A^{NP}=
-2\varepsilon^{IJKPQ}A^{PQ}\Theta^4 X^2
+2(\Theta^K\Theta^L)\varepsilon^{IJLPQ}A^{PQ}\Theta^2 X^2
\,.
\label{3.13}
\eea
The identity (\ref{X2}) follows from (\ref{XX}) while the other two are direct consequences of (\ref{A}). 

Using the identities (\ref{222}) it is not hard to verify that each line in (\ref{ids-Hi}) obeys (\ref{H-sym-1-3}). This ensures the invariance of the obtained expression for the correlation function (\ref{ans-N5}) under the interchange of operators $J^I(z_1)$ and $J^K(z_3)$.

\subsection{$\cN=5\to\cN=4$ superspace reduction}
\label{sec-reduction}

As discussed in section 1, every $\cN=5$ superconformal field theory is a special $\cN=4$ one. The $\cN=5$ supercurrent is equivalent to two $\cN=4$ 
supermultiplets, one of which is the $\cN=4$ supercurrent. 
As a result, the three-point function of the $\cN=5$ supercurrent is equivalent 
to several three-point functions in $\cN=4$ superspace. 
Here we elaborate on the $\cN=5 \to \cN=4$ superspace reduction of the 
$\cN=5$ supercurrent and its correlation functions.

We split the Grassmann coordinates $\q^\a_I $
of $\cN=5$ Minkowski superspace ${\mathbb M}^{3|10}$ onto two subsets:
(i) the coordinates $\q^\a_{\hat I}$, with $\hat I = 1, \dots, 4$,
corresponding to $\cN=4$ Minkowski superspace
${\mathbb M}^{3|8}$; and (ii) two additional coordinates $\q^\a_5$. 
The corresponding splitting of the spinor derivatives $D_\a^I$  is
$D_\a^{\hat I} $ and $D_\a^5$. Given a superfield $V$ on ${\mathbb M}^{3|10}$,
its bar-projection onto ${\mathbb M}^{3|8}$  is defined by
$V| := V|_{\q_5 =0} $.
The $\cN=5$ supercurrent $J^I$ reduces to the following 
$\cN=4$ superfields:
\bea
S^{\hat I} = J^{\hat I}|~,\qquad
J= J^5|~.
\label{3.15}
\eea
The $\cN=5$ supercurrent conservation law (\ref{J-eq}) implies 
that $S^{\hat I}$ and $J$ obey the constraints
\begin{subequations}
\bea
\label{3.16a}
D^{(\hat I}_\alpha S^{\hat J)} - \frac14 \delta^{\hat I\hat J}
D^{\hat K}_\alpha S^{\hat K} &=& 0~,\\
D^{\hat I\alpha}D^{\hat J}_\alpha J - \frac14
 \delta^{\hat I\hat J}
D^{\hat Q\alpha}D^{\hat Q}_\alpha J &=&0~.
\label{3.16b}
\eea
\end{subequations}
Eq. \eqref{3.16b} tells us that $J$ is the $\cN=4$ supercurrent \cite{BKS1,BKS2}. 
The second multiplet, $S^{\hat I}$,  
 contains among its components the
current of the fifth supersymmetry and the currents of the
remaining $\sSO(5)/\sSO(4)$ $R$-symmetry.

Note that the two-point structure $u_{12}^{IJ}$ is the integral part of two- and three-point functions (\ref{N5-two-point}) and (\ref{ans-N5}). 
Its reduction to $\cN=4$ superspace leads to
\be
\label{u-reduction}
u_{12(\cN=5)}^{\hat I\hat J}| = u_{12(\cN=4)}^{\hat I \hat J}~,
 \qquad
u_{12(\cN=5)}^{I 5}|= \delta^{I5}~.
\ee
Here we have assigned the labels $(\cN=4)$ and $(\cN=5)$ to distinguish these structures in the corresponding superspace. Below, we will omit these labels to simplify the notations. 

Using the  relations (\ref{u-reduction}) we find the $\cN=4$ superspace reduction of the two-point correlator (\ref{N5-two-point})
\begin{subequations}
\bea
\langle S^{\hat I}(z_1) S^{\hat J}(z_2) \rangle &=& c_{\cN=5}
\frac{u_{12}^{\hat I\hat J}}{{\bm x}_{12}{}^2}~,\\
\langle J(z_1) J(z_2) \rangle &=& \frac{c_{\cN=5} }{{\bm x}_{12}{}^2}
~. 
\label{N4-supercurr-2pt}
\eea
\end{subequations}
The latter correlation function coincides with the two-point correlator of the $\cN=4$ supercurrent found in \cite{BKS2} 
provided we identify  $c_{\cN=4} = c_{\cN=5}$.

The $\cN=5$ supercurrent three-point correlator (\ref{ans-N5}) reduces to the following four correlation functions of the $\cN=4$ superfields $S^{\hat I}$ and $J$
\be
\langle S^{\hat I} S^{\hat J} S^{\hat K} \rangle ~,\quad
\langle S^{\hat I} S^{\hat J} J \rangle~,\quad
\langle S^{\hat I} JJ \rangle~,\quad
\langle JJJ \rangle~,
\label{N5->4}
\ee
which can be found from different projections of the tensor (\ref{H-N5}). In particular, since $H^{\hat I\hat J\hat K}|=0$ and
$H^{\hat I 5 5}|=0$, two of the four correlation functions in (\ref{N5->4}) vanish
\be
\langle S^{\hat I} S^{\hat J} S^{\hat K} \rangle =0~,\qquad
\langle S^{\hat I} JJ \rangle =0~.
\ee
The other two correlators are non-trivial. They are 
\begin{subequations}
\bea
\langle S^{\hat I}(z_1) S^{\hat J}(z_2) J(z_3) \rangle 
&=& \frac{u_{13}^{\hat I\hat I'}
u_{23}^{\hat J\hat J'}}{{\bm x}_{13}{}^2 {\bm x}_{23}{}^2}
H_{\cN=4}^{\hat I' \hat J'}({\bm X}_3, \Theta_3)~,\\
H_{\cN=4}^{\hat I \hat J} =
H^{\hat I \hat J 5}| &=& 4d_{\cN=5} \varepsilon^{\hat I\hat J\hat K\hat L}\frac{A^{\hat K\hat L}}{X^3} + d_{\cN=5} \delta^{\hat I \hat J}\varepsilon^{\hat K\hat L\hat M\hat N}
 \frac{A^{\hat K\hat L}A^{\hat M\hat N}}{X^5}~,
\eea
\end{subequations}
and
\begin{subequations}
\bea
\langle J(z_1) J(z_2) J(z_3) \rangle &=&
\frac{H_{\cN=4}({\bm X}_3,\Theta_3)}{{\bm x}_{13}{}^2 {\bm x}_{23}{}^2 }~,\\
H_{\cN=4} = H^{555}| &=& -d_{\cN=5} \varepsilon^{\hat I\hat J\hat K\hat L}\frac{A^{\hat I\hat J}A^{\hat K\hat L}}{X^5}~.
\label{H-N4-red}
\eea
\end{subequations}

We recall that the most general form of the function $H_{\cN=4}$,
which defines the correlation function of the $\cN=4$ supercurrent,  is \cite{BKS2}
\be
H_{\cN=4} = \frac{\tilde d_{\cN=4}}{X} -d_{\cN=4} \varepsilon^{\hat I\hat J\hat K\hat L}\frac{A^{\hat I\hat J}A^{\hat K\hat L}}{X^5}~,
\label{H-N4}
\ee
where $d_{\cN=4}$ and $\tilde d_{\cN=4}$ are two independent coefficients.
Comparing (\ref{H-N4}) with (\ref{H-N4-red}) we make the following 
two conclusions: (i) only those $\cN=4$ superconformal field theories may possess extended $\cN>4$ supersymmetry for which 
\be
\tilde d_{\cN=4}=0~,
\label{tilde-d}
\ee
(ii) the coefficients $d_{\cN=4}$ and $d_{\cN=5}$ are equal,
\be
d_{\cN=4} = d_{\cN=5}~.
\label{dd}
\ee
Recall that the coefficients $c_{\cN=4}$ and $d_{\cN=4}$ are related to each other by the superconformal Ward identity \cite{BKS2}. As a consequence of (\ref{dd}) the same identity holds for $c_{\cN=5}$ and $d_{\cN=5}$:
\be
\frac{d_{\cN=4}}{c_{\cN=4}}  = \frac{d_{\cN=5}}{c_{\cN=5}}
= \frac1{16\pi}~.
\label{cd}
\ee

In conclusion of this subsection, let us briefly comment on the condition (\ref{tilde-d}) which is satisfied for all $\cN>4$ superconformal models. 
In \cite{BKS2} it was shown that the $\tilde d_{\cN=4}$-part 
of the $\cN=4$ supercurrent correlation function is non-trivial for those $\cN=4$ models which have non-equal numbers of the left and right hypermultiplets (transforming in ({\bf 2},{\bf 0}) and ({\bf 0},{\bf 2}) representations of the 
group $\sSU(2)_{\rm L}\times \sSU(2)_{\rm R}$ 
which is the double cover of the $\cN=4$ 
$R$-symmetry group $\sSO(4) \cong  \big( {\sSU}(2)_{\rm L}\times {\sSU}(2)_{\rm R} \big)/{\mathbb Z}_2$.
Thus, only those $\cN=4$ superconformal field theories may possess extended 
$\cN\geq5$ supersymmetry which contain the same number of left and right hypermultiplets.
In the next subsection we will confirm this statement by considering equations of motion of general $\cN=5$ superconformal theories. 
We will demonstrate that every  $\cN=5$ superconformal field theory 
realised 
in $\cN=4$ superspace has equal number of left and right hypermultiplets.

\subsection{Superconformal theories in $\cN=5$ superspace}
\label{sec3.1}

The $R$-symmetry group of the $\cN=5$ super-Poincar\'e algebra is 
$\sSO(5) \cong \sUSp(4)/{\mathbb Z}_2$, where the group $\sUSp(4) $
consists of matrices $g= (g_{\rm a}{}^{\rm b} )\in  \sGL(4,{\mathbb C})$ constrained by 
\bea
g^\dagger g ={\mathbbm 1}_4 ~, \qquad g^{\rm T}\O g = \O~,
\eea
for a given non-singular real symplectic metric 
$\O =(\Omega_{\rm ab}) = -\Omega^{\rm T}$.
This tensor is used to raise and lower the $\sUSp(4)$ indices,
\be
X^{\rm a} = \Omega^{\rm ab} X_{\rm b}~,\qquad
X_{\rm a} = \Omega_{\rm ab} X^{\rm b}~.
\label{Z-rule}
\ee
Here $\O^{-1} =(\Omega^{\rm ab})$ is the inverse of $\Omega$,
$\Omega_{\rm ab}\Omega^{\rm bc} = \delta_{\rm a}^{\rm c}$.

We recall that  the isomorphism $\sSO(5) \cong \sUSp(4)/{\mathbb Z}_2$
can be established by making use of a set of 
gamma-matrices
$\gamma_I = \big((\gamma_I)_{\rm a}{}^{\rm b}\big)$
for $\sSO(5)$
with the properties 
\bea
\gamma_I\gamma_J+\gamma_J \gamma_I= 2\delta_{IJ}{\mathbbm 1}_4~, 
\qquad \g_I{}^\dagger = \g_I~, \qquad \g_I{}^{\rm T} \O = \O\g_I ~, \qquad
I=1,\dots, 5~.
\eea
An explicit realisation for the matrices $\O$ and $\g_I$ is as follows:
\begin{subequations}
\bea
& & \O = \left(
\begin{array}{cc}
\ve   & 0\\
 0  &    \tilde \ve
\end{array}
\right) ~, \qquad \ve = (\ve_{ij}) =-\ve^{\rm T}~, \quad \tilde \ve 
= ( \ve_{\tilde i \tilde j} ) = -\tilde \ve{}^{\rm T}~, \quad
\ve_{12} = \ve_{\tilde 1 \tilde 2} =1
\eea
and $\g_I = (\vec{\g}, \g_4, \g_5) $, 
\bea
&&
\vec{\g}  = \left(
\begin{array}{cc}
0   & \ri \vec{\s} \\
-\ri \vec{\s}  &    0
\end{array}
\right) ~, \quad 
 \g_4 = \left(
\begin{array}{cc}
0   & {\mathbbm 1}_2\\
{\mathbbm 1}_2  &    0
\end{array}
\right) ~, \quad
 \g_5 = \left(
\begin{array}{cc}
{\mathbbm 1}_2\   & 0\\
 0  &  - {\mathbbm 1}_2\
\end{array}
\right) ~,
\eea
\end{subequations}
with $\vec \s$ the Pauli matrices. Here and below, we 
represent an $\sUSp(4)$ index as a pair $\sSU(2)$ ones, 
$X^{\rm a} = (X^i, X^{\tilde i})$.

They allow one to establish an isomorphism between ${\mathbb R}^5$
and the following linear space $\cL$ of $4\times 4$ matrices $\hat X$ constrained by 
\bea
\hat{X}{}^\dagger = \hat{X} ~, \qquad \hat{X}{}^{\rm T} \O = \O\hat{X}~, 
\qquad {\rm tr} \hat X =0~.
\eea
The isomorphism between ${\mathbb R}^5$ and $\cL$ is defined as follows: 
given a five-vector
$\vec{X} = (X^I)\in {\mathbb R}^5$, 
its image is $\hat X = X^I \g_I \in \cL$. The group $\sUSp(4)$ naturally acts 
on $\cL$ by nonsingular linear operators. Given a group element  $g \in \sUSp(4)$, 
the corresponding transformation $\hat g$ on $\cL$ is defined by 
$\hat g : \hat X \to g \hat X g^{-1}$. This induces a linear transformation $A(g)$
on 
${\mathbb R}^5$ that preserves the inner product $\langle \vec{X} | \vec{Y}\rangle 
= \frac{1}{4}{\rm tr} (\hat X \hat Y)$. It may be checked that the correspondence
$g \to A(g)$ defines a homomorphism of $\sUSp(4)$ onto $\sSO(5)$
with the kernel $ {\mathbb Z}_2 = \{ \pm {\mathbbm 1}_4\}$.

Using the symplectic metric $\O$ and its inverse $\O^{-1}$, 
let us introduce gamma-matrices with upper  and lower  indices
\be
(\gamma_I)^{\rm ab} = \Omega^{\rm ac}(\gamma_I)_{\rm c}{}^{\rm b}~,
\qquad
(\gamma_I)_{\rm ab} = \Omega_{\rm bc}(\gamma_I)_{\rm a}{}^{\rm c}
=\overline{(\gamma_I)^{\rm ab}}~.
\ee
These matrices are antisymmetric and $\Omega$-traceless
\be
(\gamma_I)^{\rm ab} = -(\gamma_I)^{\rm ba}~,\qquad
(\gamma_I)^{\rm ab} \Omega_{\rm ab} = 0~.
\ee
Thus, any $\sSO(5)$ vector $X^I$ is equivalent to
an antisymmetric $\Omega$-traceless second-rank spinor 
\be
 X^{\rm ab}:=\gamma_I^{\rm ab} X^I~,
\qquad
X^I = \frac14\gamma^I_{\rm ab}X^{\rm ab}~.
\label{USp-rule}
\ee

Using the rule (\ref{USp-rule}) we introduce the spinor covariant derivatives 
with $\sSO(5)$ spinor indices,  
$D_\a^I \to D^{\rm ab}_\alpha  = \gamma_I^{\rm ab} D^I_\alpha$. Their anti-commutation relations follow from (\ref{D-comm}),
\be
\{ D^{\rm ab}_\alpha , D^{\rm cd}_\beta \} = 2\ri (\Omega^{\rm ab}\Omega^{\rm cd}
-2\varepsilon^{\rm abcd})\partial_{\alpha\beta}~.
\ee
Let us consider a gauge theory in the $\cN=5$ superspace 
\be
\partial_{\alpha\beta} \to \nabla_{\alpha\beta}
=\partial_{\alpha\beta}+\ri V_{\alpha\beta}~,\qquad
D_\alpha^{\rm ab} \to \nabla_\alpha^{\rm ab}
=D_\alpha^{\rm ab}+\ri V_\alpha^{\rm ab}~,
\ee
where $(V_{\alpha\beta},V_\alpha^{\rm ab})$ are gauge connections.
To describe the vector multiplet, the gauge covariant derivatives 
are subject to a covariant constraint which implies
\be
\{ \nabla^{\rm ab}_\alpha , \nabla^{\rm cd}_\beta \} = 2\ri (\Omega^{\rm ab}\Omega^{\rm cd}
-2\varepsilon^{\rm abcd})\nabla_{\alpha\beta}
+\varepsilon_{\alpha\beta}(\Omega^{\rm ac}W^{\rm bd}
 - \Omega^{\rm bc}W^{\rm ad}-\Omega^{\rm ad}W^{\rm bc}+\Omega^{\rm bd}W^{\rm ac})
~.
\ee
Here $W^{\rm ab} = W^{\rm ba}$ is the field strength obeying the Bianchi identity
\be
\nabla_\alpha^{\rm a(b}W^{\rm cd)}+\frac13
\Omega^{\rm a(b}\nabla_\alpha{}^{\rm c}{}_{\rm e} W^{\rm d)e}=0~.
\label{2.366}
\ee

In complete analogy with the $\cN=6$ analysis in \cite{Samtleben},
we now consider a matter superfield $\Phi^{\rm a}$ in some representation of the gauge group. The equation of motion for the matter superfield is
\be
\nabla^{\rm a(b}_\alpha \Phi^{\rm c)} - \frac15\Omega^{\rm a(b}\nabla^{\rm c)d}_\alpha
 \Phi_{\rm d} =0~.
 \label{Phi-eq}
\ee
The consistency condition for the equation (\ref{Phi-eq}) is
\be
W^{\rm (ab} \Phi^{\rm c)} =0~.
\label{W-phi}
\ee
To solve this constraint we should assume that $W^{\rm ab}$ is a composite of the matter superfields 
\be
W^{\rm ab} = W^{\rm ab}_A T^A\,,\qquad
W^{\rm ab}_A = \ri\,\kappa\, g_{AB} \bar\Phi^{\rm (a} T^B \Phi^{\rm b)}
~,
\label{W}
\ee
where $\kappa$ is some coefficient, $T^A$ are generators of the representation and $g_{AB}$ is an invariant quadratic form on the Lie algebra of the gauge group. Note also that the Hermitian conjugate for $\Phi^{\rm a}$ is $\bar\Phi_{\rm a} = (\Phi^{\rm a})^\dag$, where we assume that $\Phi^{\rm a} = (\Phi^{\rm a}_{p})$ is a column vector in some representation of the gauge group and the letters $p,q,r,s$ from the middle of Latin alphabet denote gauge indices.

Substituting (\ref{W}) into the consistency condition (\ref{W-phi}) we find
\be
g_{AB}\bar\Phi^{p (\rm a} \Phi^{\rm b}_{q} \Phi^{\rm c)}_{s} (T^B)_{p}{}^{q}
(T^A)_{r}{}^{s} =0~,
\ee
or
\be
g_{AB}(T^A)_{p}{}^{(q} (T^B)_r{}^{s)}=0~.
\ee
The latter equation imposes strong constraints on the possible gauge group and its representations. These constraints were analysed in the works \cite{Hosomichi2008,ABJ} where the admissible gauge groups were classified.

Let us consider the $\cN=5$ supercurrent $J^I$ in the
$\sUSp(4)$ spinor notation $J^{\rm ab}=\gamma_I^{\rm ab}J^I$. The conservation law
(\ref{J-eq}) turns into
\be
D^{\rm ab}_\alpha J^{\rm cd} + D^{\rm cd}_\alpha J^{\rm ab}
+\frac1{10}(\Omega^{\rm ab}\Omega^{\rm cd} - 2\varepsilon^{\rm abcd})
D_\alpha^{\rm ef}J_{\rm ef} = 0~.
\label{J-eq1}
\ee
For the $\cN=5$ superconformal theories described by the equations (\ref{Phi-eq}) and (\ref{W}), we find the following expression for the supercurrent in terms of the matter superfields
\be
J^{\rm ab} = \bar\Phi^{\rm [a} \Phi^{\rm b]} + \frac14\Omega^{\rm ab} \bar\Phi^{\rm c} \Phi_{\rm c}~.
\ee
It is possible to check that this expression obeys (\ref{J-eq1}) due to the equations of motion (\ref{Phi-eq}) and (\ref{W}).

Now we consider the $\cN=4$ superfield reduction of $\cN=5$ superconformal models described above. This reduction amounts to setting $\theta^5_\alpha=0$ in the superfields $\Phi^{\rm a}$ and $W^{\rm ab}$. The R-symmetry group of the $\cN=4$ superspace is 
$\sSO(4) \cong  \big( {\sSU}(2)_{\rm L}\times {\sSU}(2)_{\rm R} \big)/{\mathbb Z}_2$.
This suggests that the $\sUSp(4)$ index `$\rm a$' splits into a pair of $\sSU(2)$ indices $i$ and $\tilde i$,
\be
\Phi^{\rm a} \to (q^i, q^{\tilde i})~.
\label{Phi-q}
\ee
Here $q^i$ and $q^{\tilde i}$ are left and right hypermultiplets, correspondingly. 
The $\cN=5$ gauge superfield strength $W^{\rm ab}$ has 
the following four $\cN=4$ superfield components:
\be
W^{\rm ab} \to (W^{ij}~,~ W^{i\tilde j}~,~ W^{\tilde i j}~,~ W^{\tilde i\tilde j})~.
\ee 
Here the superfields $W^{ij}$  and $W^{\tilde i\tilde j}$  constitute the field strength of 
the large $\cN=4$ vector multiplet \cite{KS15}.

 It is not hard to check that the $\cN=5$ equation (\ref{Phi-eq}) 
 leads to the standard hypermultiplet equations of motion 
\be
\nabla_\alpha^{\tilde i(j} q^{k)} =0~,\qquad
\nabla_\alpha^{i(\tilde j} q^{\tilde k)}  =0~.
\ee
The $\cN=5$ Bianchi identity \eqref{2.366}
leads to the Bianchi identities for the $\cN=4$ large vector multiplet
\be
\nabla^{\tilde i(i}_\alpha W^{jk)} = 0~,\qquad
\nabla^{i(\tilde i}_\alpha W^{\tilde j \tilde k)} = 0~. 
\ee
Reducing the equations (\ref{W}) to the $\cN=4$ superspace gives
\begin{subequations}
\bea
W^{ij}_A &=& \ri\,\kappa\, g_{AB} \bar q^{(i} T^B q^{j)}~,\qquad
W^{\tilde i\tilde j}_A = \ri\,\kappa\, g_{AB} \bar q^{(\tilde i} T^B q^{\tilde j)}~, \\
W^{i\tilde j}_A &=& \ri\,\kappa\, g_{AB} \bar q^{(i} T^B q^{\tilde
j)}~,\qquad
W^{\tilde i j}_A = \ri\,\kappa\, g_{AB} \bar q^{(\tilde i} T^B q^{j)}~.
\eea
\end{subequations}
The equations obtained coincide with the $\cN=4$ superfield equations of motion in the ABJM theory, which were given in \cite{KS15}.

The relation (\ref{Phi-q}) has the following important consequence: 
every $\cN=5$ superconformal field theory contains the same number of the left and right $\cN=4$ hypermultiplets transforming in the same representation of the gauge group. This explains the vanishing of the coefficient $\tilde d_{\cN=4}$ in (\ref{H-N4}) for those $\cN=4$ superconformal theories which possess $\cN=5$ extended supersymmetry.


\section{$\cN=6$ superconformal field theories}

Three-dimensional $\cN=6$ superconformal field theories play an important role in the $\rm AdS_4/CFT_3$ correspondence which has been intensively studied starting from the works \cite{ABJM,ABJ}. The equations of motion of such theories 
in $\cN=6$ superspace were studied in \cite{Samtleben}.
In this section we will compute the two- and three-point correlation functions of the supercurrent in these models and study their reduction to the $\cN=5$ and $\cN=4$ superspaces.

\subsection{Correlators of $\cN=6$ supercurrent}

As discussed in section 1,  the $\cN=6$ supercurrent is  a primary 
dimension-1 superfield
$J^{IJ}=-J^{JI}$ subject to the conservation equation \eqref{N6-conserv}.
The two-point function which is compatible with this conservation law is given by
\be
\label{2ptN=6}
\langle J^{IJ}(z_1) J^{KL}(z_2) \rangle
=c_{\cN=6} \frac{u_{12}^{IK} u_{12}^{JL}-u_{12}^{IL} u_{12}^{JK}}{{\bm x}_{12}{}^2}~,
\ee
where $c_{\cN=6}$ is a free coefficient.  

For the three-point correlator we make the standard ansatz
\be
\langle J^{IJ}(z_1) J^{KL}(z_2) J^{MN}(z_3) \rangle
=\frac{u_{13}^{II'} u_{13}^{JJ'}
 u_{23}^{KK'} u_{23}^{LL'} }{ {\bm x}_{13}{}^2 {\bm x}_{23}{}^2 }
H^{I'J'K'L'MN}({\bm X}_3,\Theta_3)~,
\label{3pt-N6}
\ee
where the tensor $H^{IJKLMN}=H^{[IJ][KL][MN]}$ obeys the equation
\bea
\frac23{\cal D}^P_\alpha H^{IJKLMN}
+\frac15 {\cal D}^Q_\alpha(H^{QIKLMN}\delta^{JP}
-H^{QJKLMN}\delta^{IP})\non\\
-\frac13({\cal D}^I_\alpha H^{JPKLMN}+{\cal D}^J_\alpha H^{PIKLMN})=0~,
\label{N6-H-consrev}
\eea
which is a consequence of (\ref{N6-conserv}). It has the scaling property similar to (\ref{scalingH})
\be
H^{IJKLMN}(\lambda^2 {\bm X},\lambda\Theta) = \lambda^{-2}
 H^{IJKLMN}({\bm X},\Theta)~,
\label{scalingH6}
\ee
and obeys the equations 
\bea
H^{IJKLMN}({\bm X},\Theta) &=& 
H^{KL IJ MN}(-{\bm X}^{\rm T},-\Theta)~,\label{H_1-2}\\
H^{IJKLMN}(-{\bm X}_1^{\rm T},-\Theta_1)&=&
{\bm x}_{13}{}^2 {\bm X}_3{}^2
u_{13}^{II'} u_{13}^{JJ'} u_{13}^{KK'} u_{13}^{LL'}
u_{13}^{MM'} u_{13}^{NN'}
U_3^{K'P} U_3^{L'R} 
\non\\&&\times H^{M' N' PR I'J'}(
{\bm X}_3,\Theta_3)~,
\label{H_1-3}
\eea
which follow from the invariance of the correlation function (\ref{3pt-N6}) under interchange of the order of operators.

We look for a solution of the above equations for the tensor $H^{IJKLMN}$ in the following form
\be
H^{IJKLMN}=\sum_n c_n H_n^{IJKLMN}~,
\label{H=N6}
\ee
where $c_n$ are some coefficients and 
\begin{subequations}
\label{Hn}
\bea
H^{IJKLMN}_1 &=& \frac{\varepsilon^{IJKLMN}}{X}~,\\
H^{IJKLMN}_2 &=& \frac{A^{PQ}}{X^3}[
\varepsilon^{JKMN PQ}\delta^{IL}
+\varepsilon^{ILMN PQ}\delta^{JK}
-\varepsilon^{IKMN PQ}\delta^{JL}
-\varepsilon^{JLMN PQ}\delta^{IK}]~,\\
H^{IJKLMN}_3 &=& \frac{A^{PQ}}{X^3}[\varepsilon^{IJ KM PQ}\delta^{LN}
-\varepsilon^{IJ LM PQ}\delta^{KN}
-\varepsilon^{IJ KN PQ}\delta^{LM}
+\varepsilon^{IJ LN PQ}\delta^{KM}]~,\\
H^{IJKLMN}_4 &=&\frac{A^{PQ}}{X^3}[
\varepsilon^{KL JM PQ}\delta^{IN}
+\varepsilon^{KL IN PQ}\delta^{JM}
-\varepsilon^{KL IM PQ}\delta^{JN}
-\varepsilon^{KL JN PQ}\delta^{IM}]~,\\
H_5^{IJKLMN} &=&\frac{A^{PQ}A^{RS}}{X^5}
[\varepsilon^{MN PQRS}(\delta^{JK}\delta^{IL}-\delta^{IK}\delta^{JL})]~,\\
H_6^{IJKLMN} &=&\frac{A^{PQ}A^{RS}}{X^5}
[\varepsilon^{IJ PQRS}(\delta^{LM}\delta^{KN}-\delta^{KM}\delta^{LN})]~,\\
H_7^{IJKLMN} &=&\frac{A^{PQ}A^{RS}}{X^5}
[\varepsilon^{KL PQRS}(\delta^{JM}\delta^{IN}-\delta^{IM}\delta^{JN})]~,\\
H_8^{IJKLMN} &=&\frac{A^{PQ}A^{RS}}{X^5}
[\varepsilon^{JK PQRS}(\delta^{IM}\delta^{LN}-\delta^{IN}\delta^{LM})-\varepsilon^{IK PQRS}(\delta^{JM}\delta^{LN}-\delta^{JN}\delta^{LM})
\non\\&&
-\varepsilon^{JL PQRS}(\delta^{IM}\delta^{KN}-\delta^{IN}\delta^{KM})+\varepsilon^{IL PQRS}(\delta^{JM}\delta^{KN}-\delta^{JN}\delta^{KM})]~,\\
H_9^{IJKLMN} &=&\frac{A^{PQ}A^{RS}}{X^5}
[\varepsilon^{IM PQRS}(\delta^{JK}\delta^{LN}-\delta^{JL}\delta^{KN})
-\varepsilon^{JM PQRS}(\delta^{IK}\delta^{LN}-\delta^{IL}\delta^{KN})
\non\\&&
-\varepsilon^{IN PQRS}(\delta^{JK}\delta^{LM}-\delta^{JL}\delta^{KM})
+\varepsilon^{JN PQRS}(\delta^{IK}\delta^{LM}-\delta^{IL}\delta^{KM})]~,\\
H_{10}^{IJKLMN} &=&\frac{A^{PQ}A^{RS}}{X^5}
[\varepsilon^{KM PQRS}(\delta^{IL}\delta^{JN}-\delta^{JL}\delta^{IN})
-\varepsilon^{LM PQRS}(\delta^{IK}\delta^{JN}-\delta^{JK}\delta^{IN})
\non\\&&
-\varepsilon^{KN PQRS}(\delta^{IL}\delta^{JM}-\delta^{JL}\delta^{IM})
+\varepsilon^{LN PQRS}(\delta^{IK}\delta^{JM}-\delta^{JK}\delta^{IM})]~,\\
H_{11}^{IJKLMN}&=&\frac{\varepsilon^{PQRSTU}A^{PQ}A^{RS}A^{TU}
}{X^7}[\delta^{JK}(\delta^{IM}\delta^{LN}-\delta^{IN}\delta^{LM})
-\delta^{IK}(\delta^{JM}\delta^{LN}-\delta^{JN}\delta^{LM})
\non\\&&+\delta^{IL}(\delta^{JM}\delta^{KN}-\delta^{JN}\delta^{KM})
-\delta^{JL}(\delta^{IM}\delta^{KN}-\delta^{IN}\delta^{KM})
]~.
\eea
\end{subequations}
The tensors (\ref{Hn}) obey the constraints (\ref{scalingH6}) and (\ref{H_1-2}) by construction. Imposing the equation (\ref{N6-H-consrev}) we find the coefficients $c_n$
\be
\begin{array}c
c_1=2d_{\cN=6}~,\quad
c_2=c_3=c_4 = d_{\cN=6}~,\quad
c_{11}=\frac1{24}d_{\cN=6}~,\\
c_5=c_6=c_7=c_8=c_9=c_{10} = \frac14 d_{\cN=6}~,
\end{array}
\ee
where $d_{\cN=6}$ is a free coefficient.
It is possible to show that for these values of the coefficients the tensor (\ref{H=N6}) obeys the equation (\ref{H_1-3}) which ensures the invariance of the correlation function under the interchange of operators $J^{IJ}(z_1)$ and $J^{MN}(z_3)$. However, it is a tedious exercise  to demonstrate this directly.
Instead of embarking on such an exercise, we will take a shortcut 
 and prove the required 
 symmetry property using the $\cN=6 \to \cN=5$ superspace reduction 
 of the superfield operators and their correlation functions.

\subsection{$\cN=6\to\cN=5$ superspace reduction}
Let us split the $\sSO(6)$ index $I$ as $I=(\hat I,6)$,
$\hat I=1,2,3,4,5$. Upon reduction to $\cN=5$ superspace, 
the $\cN=6$ supercurrent $J^{IJ}$ leads to two $\cN=5$ superfields, 
one of which is 
the $\cN=5$ supercurrent $J^{\hat I}$ and the other is an antisymmetric 
tensor $K^{\hat I\hat J}$:
\be
J^{\hat I} = J^{\hat I 6}|~,\qquad
K^{\hat I\hat J} = J^{\hat I \hat J}|~, 
\ee
where the bar-projection means setting $\theta^6_\alpha=0$. 
As a consequence of the $\cN=6$ supercurrent conservation 
equation (\ref{N6-conserv}),  
$J^{\hat I}$ proves to obey the   $\cN=5$ supercurrent conservation equation 
(\ref{J-eq}), while for $K^{\hat I\hat J}$ we obtain the following constraint
\be
D^{\hat I}_\alpha K^{\hat J \hat K}
= D^{[\hat I}_\alpha K^{\hat J \hat K]}
-\frac12 D^{\hat L}_\alpha K^{\hat L[\hat J}
 \delta^{\hat K]\hat I}~.
\label{4.12}
\ee

The $\cN=6\to\cN=5$ superspace reduction of the two-point structure $u_{12}^{IJ}$ is similar to (\ref{u-reduction}). Therefore, the reduction of the two-point correlation function (\ref{2ptN=6}) is rather trivial
\begin{subequations}
\bea
\langle K^{\hat I \hat J}(z_1) K^{\hat K \hat L}(z_2) \rangle
 &=& c_{\cN=6} \frac{u_{12}^{\hat I\hat K} u_{12}^{\hat J\hat L}-u_{12}^{\hat I\hat L} u_{12}^{\hat J\hat K}}{{\bm x}_{12}{}^2}~,
 \\
\langle J^{\hat I}(z_1) J^{\hat J}(z_2) \rangle
&=&c_{\cN=6} \frac{u_{12}^{\hat I\hat J}}{{\bm x}_{12}{}^2}~.
\eea
\end{subequations}

The three-point function (\ref{3pt-N6}) reduces to the following four correlation functions in the $\cN=5$ superspace
\be
\langle K^{\hat I\hat J} K^{\hat K\hat L} K^{\hat M\hat N} \rangle ~,\quad
\langle K^{\hat I\hat J} K^{\hat K\hat L} J^{\hat M} \rangle ~,\quad
\langle K^{\hat I\hat J} J^{\hat K} J^{\hat L} \rangle ~,\quad
\langle J^{\hat I} J^{\hat J} J^{\hat K} \rangle ~.
\ee
These correlators can be obtained by considering different components of the tensor (\ref{H=N6}). In particular, from the explicit form of the tensors (\ref{Hn}) we immediately see that two of the four correlators vanish
\be
\langle K^{\hat I\hat J} K^{\hat K\hat L} K^{\hat M\hat N} \rangle=0~,\qquad
\langle K^{\hat I\hat J} J^{\hat K} J^{\hat L} \rangle =0~.
\ee
For the other two we find
\begin{subequations}
\label{4.15}
\bea
\langle J^{\hat I}(z_1) J^{\hat J}(z_2) J^{\hat K}(z_3) \rangle
&=&\frac{u_{13}^{\hat I\hat I'} u_{23}^{\hat J\hat J'}}{
{\bm x}_{13}{}^2 {\bm x}_{23}{}^2} 
 H^{\hat I'\hat J'\hat K}_{\cN=5}({\bm X}_3,\Theta_3)~,\\
 H^{\hat I\hat J\hat K}_{\cN=5} =H^{\hat I6 \hat J6 \hat K6}|&=&
d_{\cN=6}\frac14 
(\delta^{\hat I\hat J}\varepsilon^{\hat K\hat L\hat M\hat N\hat P} 
-\delta^{\hat I\hat K} \varepsilon^{\hat J\hat L\hat M\hat N\hat P}
-\delta^{\hat J\hat K} \varepsilon^{\hat I\hat L\hat M\hat N\hat P})
\frac{A^{\hat L\hat M}A^{\hat N\hat P}}{X^5}
\non\\&&+d_{\cN=6} \varepsilon^{\hat I\hat J\hat K\hat L\hat M}\frac{A^{\hat L\hat M}}{X^3}~,
\label{4.15b}
\eea
\end{subequations}
and
\begin{subequations}
\label{4.16}
\bea
&&\langle K^{\hat I\hat J}(z_1) J^{\hat K}(z_2) K^{\hat M\hat N}(z_3) \rangle
= \frac{u_{13}^{\hat I\hat I'}u_{13}^{\hat J\hat J'}u_{23}^{\hat K\hat K'}}{{\bm x}_{13}{}^2
 {\bm x}_{23}{}^2} H_{\cN=5}^{\hat I'\hat J'\hat K'\hat M\hat N}({\bm X}_3,\Theta_3)~,\label{LJL}\\
&&
 H_{\cN=5}^{\hat I\hat J\hat K\hat M\hat N}=H^{\hat I\hat J\hat K6\hat M\hat N}|=2d_{\cN=6}\frac{\varepsilon^{\hat I\hat J\hat K\hat M\hat N}}{X}
\non\\&&
+d_{\cN=6}\frac{A^{\hat P\hat Q}}{X^3}[
\varepsilon^{\hat I\hat M\hat N\hat P\hat Q}\delta^{\hat J\hat K}
-\varepsilon^{\hat J\hat M\hat N\hat P\hat Q}\delta^{\hat I\hat K}
-\varepsilon^{\hat I\hat J\hat N\hat P\hat Q}\delta^{\hat K\hat M}\non\\&&
+\varepsilon^{\hat I\hat J\hat M\hat P\hat Q}\delta^{\hat K\hat N}
-\varepsilon^{\hat K\hat I\hat M\hat P\hat Q}\delta^{\hat J\hat N}
+\varepsilon^{\hat K\hat J\hat M\hat P\hat Q}\delta^{\hat I\hat N}
+\varepsilon^{\hat K\hat I\hat N\hat P\hat Q}\delta^{\hat J\hat M}
-\varepsilon^{\hat K\hat J\hat N\hat P\hat Q}\delta^{\hat I\hat M}
]\non\\&&
+\frac14d_{\cN=6}\frac{A^{\hat P\hat Q}A^{\hat R\hat S}}{X^5}
[\varepsilon^{\hat K\hat P\hat Q\hat R\hat S}(\delta^{\hat J\hat M}\delta^{\hat I\hat N}-\delta^{\hat I\hat M}\delta^{\hat J\hat N})
-\varepsilon^{\hat J\hat P\hat Q\hat R\hat S}(\delta^{\hat I\hat M}\delta^{\hat K\hat N}-\delta^{\hat K\hat M}\delta^{\hat I\hat N})\non\\&&
+\varepsilon^{\hat I\hat P\hat Q\hat R\hat S}(\delta^{\hat J\hat M}\delta^{\hat K\hat N}-\delta^{\hat J\hat N}\delta^{\hat K\hat M})
+\varepsilon^{\hat M\hat P\hat Q\hat R\hat S}(\delta^{\hat I\hat K}\delta^{\hat J\hat N}-\delta^{\hat J\hat K}\delta^{\hat I\hat N})
\non\\&&
-\varepsilon^{\hat N\hat P\hat Q\hat R\hat S}(\delta^{\hat I\hat K}\delta^{\hat J\hat M}-\delta^{\hat J\hat K}\delta^{\hat I\hat M})
]~.
\label{4.16b}
\eea
\end{subequations}
Comparing (\ref{4.15b}) with (\ref{H-N5}) we find the relation among the coefficients
\be
d_{\cN=6} = 4 d_{\cN=5}~.
\ee
This relation, in conjunction with (\ref{cd}), gives us the ratio between the coefficients of two- and three-point functions
\be
\frac{d_{\cN=6}}{c_{\cN=6}} = 4 \frac{d_{\cN=5}}{c_{\cN=5}}
=\frac1{4\pi}~.
\ee
This relation is a manifestation of a Ward identity which relates the supercurrent two-point and three-point functions.

The tensor (\ref{4.16b}) can be represented in the equivalent form
\begin{subequations}
\label{4.21}
\be
H_{\cN=5}^{\hat I \hat J \hat K \hat L \hat M}=d_{\cN=6}\sum_n a_n  H_n^{\hat I\hat J\hat K\hat L\hat M}~,
\label{H-cov}
\ee
where the coefficients $a_n$ are 
\be
\begin{array}c
a_1 = a_2 =1~, \quad
a_3 = a_4 =-\frac12~, \quad
a_5 = a_6 = a_7 = a_8 = \frac12~,\quad
a_9 = -\frac14~,\\
a_{10} = a_{11} = a_{12} = a_{13} = a_{14} = a_{15}=\frac18~,\quad
a_{16} = -\frac38~,\quad 
a_{17} = \frac3{64}~,
\end{array}
\ee
and the tensors $ H^{\hat I\hat J\hat K\hat L\hat M}_n$ are expressed in terms of the covariant objects  (\ref{three-points})
\bea
 H_1^{\hat I\hat J\hat K\hat M\hat N}&=&\frac{\varepsilon^{\hat I\hat J\hat K\hat M\hat N} }{\bm X}~,\\
 H_2^{\hat I\hat J\hat K\hat M\hat N}&=&\frac{U^{\hat K'\hat K}\varepsilon^{\hat I\hat J\hat K'\hat M\hat N}}{\bm X}~,\\
 H_3^{\hat I\hat J\hat K\hat M\hat N}&=&-\frac12\frac{U^{\hat P\hat Q}}{\bm X}
 (\varepsilon^{\hat J\hat M\hat N\hat P\hat Q}\delta^{\hat I\hat K}-\varepsilon^{\hat I\hat M\hat N\hat P\hat Q}\delta^{\hat J\hat K})
~,\\
 H_4^{\hat I\hat J\hat K\hat M\hat N}&=&\frac12\frac{U^{\hat P\hat Q}}{\bm X}
 (U^{\hat M\hat K}\varepsilon^{\hat N\hat I\hat J\hat P\hat Q}-U^{\hat N\hat K}\varepsilon^{\hat M\hat I\hat J\hat P\hat Q})
 ~,\\
 H_5^{\hat I\hat J\hat K\hat M\hat N}&=&-\frac12\frac{U^{\hat P\hat Q}}{\bm X}
 (\varepsilon^{\hat I\hat J\hat M\hat P\hat Q}\delta^{\hat K\hat N}-\varepsilon^{\hat I\hat J\hat N\hat P\hat Q}\delta^{\hat K\hat M})
~,\\
 H_6^{\hat I\hat J\hat K\hat M\hat N}&=&\frac12\frac{U^{\hat P\hat Q}}{\bm X}
(\varepsilon^{\hat K\hat I\hat M\hat P\hat Q}\delta^{\hat J\hat N}
 - \varepsilon^{\hat K\hat J\hat M\hat P\hat Q}\delta^{\hat I\hat N}
 - \varepsilon^{\hat K\hat I\hat N\hat P\hat Q}\delta^{\hat J\hat M}
 + \varepsilon^{\hat K\hat J\hat N\hat P\hat Q}\delta^{\hat I\hat M})~,\\
 H_7^{\hat I\hat J\hat K\hat M\hat N}&=&\frac12\frac{U^{\hat P\hat Q}}{\bm X}
(U^{\hat J\hat K}\varepsilon^{\hat M\hat N\hat I\hat P\hat Q}-U^{\hat I\hat K}\varepsilon^{\hat M\hat N\hat J\hat P\hat Q})~,\\
 H_8^{\hat I\hat J\hat K\hat M\hat N}&=&\frac12\frac{U^{\hat P\hat Q}U^{\hat L\hat K}}{\bm X}
(\delta^{\hat J\hat M}\varepsilon^{\hat L\hat N\hat I\hat P\hat Q}+\delta^{\hat I\hat N}\varepsilon^{\hat L\hat M\hat J\hat P\hat Q}-\delta^{\hat J\hat N}\varepsilon^{\hat L\hat M\hat I\hat P\hat Q}-\delta^{\hat I\hat M}\varepsilon^{\hat L\hat N\hat J\hat P\hat Q})
~,\\
 H_{9}^{\hat I\hat J\hat K\hat M\hat N}&=&\frac14\frac{U^{\hat P\hat Q}U^{\hat R\hat S}}{\bm X}
\varepsilon^{\hat K\hat P\hat Q\hat R\hat S}(\delta^{\hat I\hat M}\delta^{\hat J\hat N}-\delta^{\hat J\hat M}\delta^{\hat I\hat N})
~,\\
 H_{10}^{\hat I\hat J\hat K\hat M\hat N}&=&\frac14\frac{U^{\hat P\hat Q}U^{\hat R\hat S}}{\bm X}
[\varepsilon^{\hat I\hat P\hat Q\hat R\hat S}(\delta^{\hat J\hat M}\delta^{\hat K\hat N}-\delta^{\hat J\hat N}\delta^{\hat K\hat M})
-(\hat I\leftrightarrow \hat J)]~,\\
 H_{11}^{\hat I\hat J\hat K\hat M\hat N}&=&\frac14\frac{U^{\hat P\hat Q}U^{\hat R\hat S}}{\bm X}
[\varepsilon^{\hat M\hat P\hat Q\hat R\hat S}(\delta^{\hat I\hat N}U^{\hat J\hat K}-\delta^{\hat J\hat N}U^{\hat I\hat K})
-(\hat M \leftrightarrow \hat N)]~,\\
 H_{12}^{\hat I\hat J\hat K\hat M\hat N}&=&\frac14\frac{U^{\hat P\hat Q}U^{\hat R\hat S}}{\bm X}
[\varepsilon^{\hat M\hat P\hat Q\hat R\hat S}(\delta^{\hat I\hat K}\delta^{\hat J\hat N}-\delta^{\hat J\hat K}\delta^{\hat I\hat N})
-(\hat M \leftrightarrow \hat N)]~,\\
 H_{13}^{\hat I\hat J\hat K\hat M\hat N}&=&\frac14\frac{U^{\hat P\hat Q}U^{\hat R\hat S}}{\bm X}
[\varepsilon^{\hat J\hat P\hat Q\hat R\hat S}(\delta^{\hat I\hat M}U^{\hat N\hat K}-\delta^{\hat I\hat N}U^{\hat M\hat K})
-(\hat I \leftrightarrow \hat J)]~,\\
 H_{14}^{\hat I\hat J\hat K\hat M\hat N}&=&\varepsilon^{\hat I\hat J\hat K\hat M\hat N}\frac{\Theta^4}{{\bm X}^3}~,\\
 H_{15}^{\hat I\hat J\hat K\hat M\hat N}&=&\frac{U^{\hat K'\hat K}\varepsilon^{\hat I\hat J\hat K'\hat M\hat N}\Theta^4}{{\bm X}^3}~,\\
 H_{16}^{\hat I\hat J\hat K\hat M\hat N}&=&-\frac12\frac{\Theta^4}{{\bm X}^3}
 U^{\hat P\hat Q}(\varepsilon^{\hat K\hat I\hat M\hat P\hat Q}\delta^{\hat J\hat N}
  - \varepsilon^{\hat K\hat J\hat M\hat P\hat Q}\delta^{\hat I\hat N}
  - \varepsilon^{\hat K\hat I\hat N\hat P\hat Q}\delta^{\hat J\hat M}
  + \varepsilon^{\hat K\hat J\hat N\hat P\hat Q}\delta^{\hat I\hat M})~,~~~~~\\
 H_{17}^{\hat I\hat J\hat K\hat M\hat N}&=&\frac{\Theta^8}{{\bm X}^5}\varepsilon^{\hat I\hat J\hat K\hat M\hat N}~.
\eea
\end{subequations}
In verifying that (\ref{4.21}) coincides with (\ref{4.16b}),
it is advantageous to use the relations (\ref{X2})--(\ref{3.13}) as well as the following $\cN=5$ superspace identities 
\bea
A^{IJ}A^{KL}&=&A^{[IJ}A^{KL]}
-\frac13X^2(B^{IK}B^{JL}-B^{JK}B^{IL})~,\\
\frac43B^{QK}(B^{PI}\varepsilon^{MNJPQ}-B^{PJ}\varepsilon^{MNIPQ})
&=&\frac43B^{PK}(B^{QM}\varepsilon^{IJNPQ}-B^{QN}\varepsilon^{IJMPQ})
\non\\&&-2\Theta^2 B^{PK}\varepsilon^{IJMNP}~,\\
A^{PQ}(B^{IK}\varepsilon^{MNJPQ}-B^{JK}\varepsilon^{MNIPQ})
&=&A^{PQ}(B^{NK}\varepsilon^{IJPQM}-B^{MK}\varepsilon^{IJNPQ})
\non\\&&-\Theta^2 A^{PK}\varepsilon^{IJMNP}~,
\eea
where $B^{IJ}\equiv \Theta^{I\alpha}\Theta^J_\alpha$.

For the tensor $ H^{\hat I\hat J\hat K\hat M\hat N}$ in the form 
(\ref{4.21}) it is possible to check that it obeys the equation
\be 
H^{\hat I\hat J\hat K\hat M\hat N}(-{\bm X}_1^{\rm T},-\Theta_1)=
{\bm x}_{13}{}^2 {\bm X}_3{}^2
u_{13}^{\hat I\hat I'} u_{13}^{\hat J\hat J'} u_{13}^{\hat K\hat K'} 
u_{13}^{\hat M\hat M'} u_{13}^{\hat N\hat N'}
U_3^{\hat K'\hat P}  H^{\hat M' \hat N' \hat P \hat I'\hat J'}(
{\bm X}_3,\Theta_3)~,
\ee
which ensures the invariance of the correlation function (\ref{LJL}) under the interchange of operators $K^{\hat I\hat J}(z_1)$ and $K^{\hat M\hat N}(z_3)$. Thus, the two $\cN=5$ projections (\ref{4.15}) and (\ref{4.16}) are invariant under interchange of positions of operators. This proves that the $\cN=6$ correlation function (\ref{3pt-N6}) also respects this symmetry.


\subsection{Further $\cN=5\to\cN=4$ superspace reduction}

The $\cN=6$ correlator (\ref{3pt-N6}) reduces to two $\cN=5$ correlation functions (\ref{4.15}) and (\ref{4.16}), one of which is just the $\cN=5$ supercurrent correlator while the other is the mixed correlator. The reduction to the $\cN=4$ superspace of the $\cN=5$ supercurrent correlation function was considered in section \ref{sec-reduction}. Here we will study the $\cN=4$ superspace reduction of the mixed correlator (\ref{4.16}). 

In this subsection the $\sSO(5)$ indices are denoted by $I,J,K,\ldots$, while the $\sSO(4)$ indices are denoted by the same letters with hats, e.g., $\hat I,\hat J,\hat K,\ldots$ The $\cN=4$ superspace components of the $\cN=5$ supercurrent are given in (\ref{3.15}). We define the $\cN=4$ superspace projections of the antisymmetric tensor $K^{IJ}$  as
\be
L^{\hat I \hat J} = K^{\hat I\hat J}| ~,\qquad
R^{\hat I} = K^{\hat I 5}|~,
\ee
where the bar-projection means $\theta^5_\alpha =0$.
As a consequence of (\ref{4.12}) they obey the following equations
\bea
&&D^{\hat I}_\alpha L^{\hat J\hat K} =
D^{[\hat I}_\alpha L^{\hat J\hat K]} - \frac23
D^{\hat L}_\alpha L^{\hat L[\hat J}\delta^{\hat K]\hat I}~,\\
&&D^{(\hat I}_\alpha R^{\hat J)}-\frac14 \delta^{\hat I\hat J}
D^{\hat K}_\alpha R^{\hat K} =0~.
\eea
The latter equation coincides with (\ref{3.16a}) while the former shows that $L^{\hat I\hat J}$ describes the $\cN=4$ flavour current multiplets. The antisymmetric tensor $L^{\hat I\hat J}$ can be further decomposed into two components $L_+^{\hat I\hat J}$ and $L^{\hat I\hat J}_-$ with different self-duality properties
\be
\frac12 \varepsilon^{\hat I\hat J\hat K\hat L} L_\pm^{\hat K\hat L} = \pm L_\pm^{\hat I\hat J}~.
\ee

The correlator $\langle K^{IJ} J^K K^{MN} \rangle $ reduces to the following six correlation functions of the $\cN=4$ superfields
\be
\langle L^{\hat I\hat J} S^{\hat K} L^{\hat M\hat N} \rangle~,\quad
\langle L^{\hat I\hat J} J L^{\hat M\hat N} \rangle~,\quad
\langle L^{\hat I\hat J} S^{\hat K} R^{\hat M} \rangle~, \quad
\langle L^{\hat I\hat J} J R^{\hat M} \rangle~,\quad
\langle R^{\hat I} S^{\hat K} R^{\hat M} \rangle~, \quad
\langle R^{\hat I} J R^{\hat M} \rangle~.
\ee 
They can be found by taking $\cN=4$ superspace projections of the tensor (\ref{4.16b}). In particular, it is easy to see that three of these six correlators vanish
\be
\langle L^{\hat I\hat J} S^{\hat K} L^{\hat M\hat N} \rangle=0~,
\quad
\langle L^{\hat I\hat J} J R^{\hat M} \rangle=0~,\quad
\langle R^{\hat I} S^{\hat K} R^{\hat M} \rangle=0~.
\ee
For the other three we find
\begin{subequations}
\label{LJL}
\bea
\langle L^{\hat I\hat J}(z_1) J(z_2) L^{\hat M\hat N}(z_3) \rangle
 &=& \frac{u_{13}^{\hat I\hat I'}u_{13}^{\hat J\hat J'}}{
 {\bm x}_{13}{}^2 {\bm x}_{23}{}^2}H^{\hat I'\hat J' \hat M\hat N}({\bm X}_3,\Theta_3) ~,\\
 H^{\hat I\hat J \hat M\hat N}&=&d_{\cN=6}\bigg[2 \frac{\varepsilon^{\hat I\hat J\hat M\hat N}}{X}
 +\frac14\frac{A^{\hat P\hat Q}A^{\hat R\hat S}}{X^5}
 \varepsilon^{\hat P\hat Q\hat R\hat S}(
  \delta^{\hat J\hat M}\delta^{\hat I\hat N}-\delta^{\hat I\hat M}\delta^{\hat J\hat N})\non\\&+&
 \frac{A^{\hat P\hat Q}}{X^3}(
 \varepsilon^{\hat J\hat M\hat P\hat Q}\delta^{\hat I\hat N}
 +\varepsilon^{\hat I\hat N\hat P\hat Q}\delta^{\hat J\hat M}
 -\varepsilon^{\hat J\hat N\hat P\hat Q}\delta^{\hat I\hat M}
 -\varepsilon^{\hat I\hat M\hat P\hat Q}\delta^{\hat J\hat N})\bigg]~,~~~~~~~~
\eea
\end{subequations}

\begin{subequations}
\bea
&&\langle L^{\hat I\hat J}(z_1) S^{\hat K}(z_2) R^{\hat L}(z_3) \rangle =\frac{u_{13}^{\hat I\hat I'}u_{13}^{\hat J\hat J'}
 u_{23}^{\hat K\hat K'}}{
 {\bm x}_{13}{}^2 {\bm x}_{23}{}^2}H^{\hat I'\hat J' \hat K'\hat L}({\bm X}_3,\Theta_3)~,\\
&&H^{\hat I\hat J \hat K\hat L}=
d_{\cN=6}\bigg[2 \frac{\varepsilon^{\hat I\hat J\hat K\hat L}}{X}
 +\frac14\frac{A^{\hat P\hat Q}A^{\hat R\hat S}}{X^5}
 \varepsilon^{\hat P\hat Q\hat R\hat S}(
  \delta^{\hat J\hat K}\delta^{\hat I\hat L}-\delta^{\hat I\hat K}\delta^{\hat J\hat L})
 \non\\&&\qquad\qquad
 + \frac{A^{\hat P\hat Q}}{X^3}(
 \varepsilon^{\hat I\hat L\hat P\hat Q}\delta^{\hat J\hat K}
 -\varepsilon^{\hat J\hat L\hat P\hat Q}\delta^{\hat I\hat K}
 -\varepsilon^{\hat I\hat J\hat P\hat Q}\delta^{\hat K\hat L}
 +\varepsilon^{\hat K\hat I\hat P\hat Q}\delta^{\hat J\hat L}
 -\varepsilon^{\hat K\hat J\hat P\hat Q}\delta^{\hat I\hat L})
 \bigg]~,~~~~~~~~
\eea
\end{subequations}

\begin{subequations}
\bea
\langle R^{\hat I}(z_1) J(z_2) R^{\hat J}(z_3) \rangle
&=&\frac{u_{13}^{\hat I\hat I'}}{{\bm x}_{13}{}^2 {\bm x}_{23}{}^2}
H^{\hat I' \hat J}({\bm X}_3,\Theta_3)~,\\
H^{\hat I \hat J}&=&-d_{\cN=6}\left[
\varepsilon^{\hat I\hat J\hat K\hat L}\frac{A^{\hat K\hat L}}{X^3}
+\frac14\delta^{\hat I\hat J}\varepsilon^{\hat K\hat L\hat M\hat N}\frac{A^{\hat K\hat L}A^{\hat M\hat N}}{X^5}
\right]~.
\eea
\end{subequations}

The correlation function (\ref{LJL}) can be decomposed into two parts with opposite self-duality properties
\begin{subequations}
\bea
\langle L_\pm^{\hat I\hat J}(z_1) J(z_2) L_\pm^{\hat M\hat N}(z_3) \rangle
 &=& \frac{u_{13}^{\hat I\hat I'}u_{13}^{\hat J\hat J'}}{
 {\bm x}_{13}{}^2 {\bm x}_{23}{}^2}H_\pm^{\hat I'\hat J' \hat M\hat N}({\bm X}_3,\Theta_3) ~,\label{4.32a}\\
H_\pm^{\hat I\hat J\hat M\hat N}  &=&\frac12H^{\hat I\hat J\hat M\hat N}\pm\frac14\varepsilon^{\hat I\hat J\hat K\hat L}H^{\hat K\hat L\hat M\hat N}
=d_{\cN=6}\frac1X(\varepsilon^{\hat I\hat J\hat M\hat N}\pm\delta^{\hat I\hat M}\delta^{\hat J\hat N}\mp\delta^{\hat I\hat N}\delta^{\hat J\hat M})
\non\\&&-\frac18d_{\cN=6}\frac{A^{\hat P\hat Q}A^{\hat R\hat S}}{X^5} \varepsilon^{\hat P\hat Q\hat R\hat S}
(\delta^{\hat I\hat M}\delta^{\hat J\hat N}-\delta^{\hat J\hat M}\delta^{\hat I\hat N}\pm\varepsilon^{\hat I\hat J\hat M\hat N})
\non\\&&
+\frac12d_{\cN=6}\frac{A^{\hat P\hat Q}}{X^3}[\varepsilon^{\hat J\hat M\hat P\hat Q}\delta^{\hat I\hat N}
+\varepsilon^{\hat I\hat N\hat P\hat Q}\delta^{\hat J\hat M}-\varepsilon^{\hat J\hat N\hat P\hat Q}\delta^{\hat I\hat M}
 - \varepsilon^{\hat I\hat M\hat P\hat Q}\delta^{\hat J\hat N}
 \non\\&&
 \pm2\delta^{\hat P\hat J}(\delta^{\hat N\hat I}\delta^{\hat Q\hat M}-\delta^{\hat M\hat I}\delta^{\hat Q\hat N})
 \mp2\delta^{\hat P\hat I}(\delta^{\hat N\hat J}\delta^{\hat Q\hat M}-\delta^{\hat M\hat J}\delta^{\hat Q\hat M})
 ]~.
\eea
\end{subequations}
This correlator was found in \cite{BKS2} in another form within the iso-spinor formalism. We point out that in (\ref{4.32a}) there is no mixed correlator involving both $L_+^{\hat I\hat J}$ and $L_-
^{\hat I\hat J}$.

\section{Discussion}

As continuation of the program initiated in \cite{BKS1,BKS2}, 
in this paper we have computed 
the two- and three-point correlation functions of the supercurrent multiplets 
in general $\cN=5,6$ superconformal field theories in three dimensions.
We demonstrated that the functional form of each of these correlators is 
completely determined  by the superconformal symmetry modulo a single overall coefficient. The ratio of the coefficients arising in the two-point and three-point functions is fixed by the Ward identities. The remaining coefficients are model-dependent. 

Every $\cN=5$ or $\cN=6$ superconformal field theory can be viewed 
as a special $\cN=4$ superconformal field theory. 
We demonstrated that the general property of the
 $\cN=5$ or $\cN=6$ superconformal field theories 
is that they are invariant under the $\cN=4$ mirror map. 

As is explained in section 1, the $\cN=5$ supercurrent is described by an iso-vector $J^I$ while the $\cN=6$ supercurrent is given by an antisymmetric tensor $J^{IJ}$. As a consequence, their three-point correlation functions are specified by rank-3 $H^{IJK}$ and rank-6 $H^{IJKLMN}$ tensors, respectively.
Although the form of the tensor $H^{IJK}$ is relatively compact, see (\ref{H-N5}), the $\cN=6$ tensor $H^{IJKLMN}$ has rather  clumsy form because of proliferation of $\sSO(6)$ indices (\ref{Hn}). It is desirable to develop a superspace formalism 
that provides a compact form for these correlators. It is natural to expect that this should be a version of harmonic/projective superspace since supercurrents 
in such superspaces may be realised as scalar superfields. 

A few years ago, Ref. \cite{KPT-MvU} presented a family of 
homogeneous spaces, $\overline{\mathbb M}{}^{3|2\cN} \times {\mathbb X}^\cN_m$, 
of the 3D $\cN$-extended superconformal group 
$\sOSp(\cN|2, {\mathbb R})$,  for any positive integer
$m\leq [\cN/2]$, with  $[\cN/2]$
the integer part of $\cN/2$.   Here $\overline{\mathbb M}{}^{3|2\cN}$ 
denotes  the compactified $\cN$-extended Minkowski superspace on which 
the superconformal group $\sOSp(\cN|2, {\mathbb R})$ acts by well-defined 
transformations. The usual Minkowski superspace is embedded in 
$\overline{\mathbb M}{}^{3|2\cN}$ as a dense open subset. 
The internal sector $ {\mathbb X}^\cN_m$
of $\overline{\mathbb M}{}^{3|2\cN} \times {\mathbb X}^\cN_m$
is realised in terms of odd supertwistors  subject to certain conditions \cite{KPT-MvU}. 
For many applications, it suffices to work with 
the dense open subset ${\mathbb M}{}^{3|2\cN} \times {\mathbb X}^\cN_m$ 
of $\overline{\mathbb M}{}^{3|2\cN} \times {\mathbb X}^\cN_m$. 
Then the points of $ {\mathbb X}^\cN_m$ can be identified with 
$m$ complex $\cN$-vectors 
$Z^{\ju} =(Z_I{}^{\ju}) \in {\mathbb C}^\cN -\{ 0\} $ which are required to
(i) be linearly 
independent;  (ii) obey the null conditions
\bea
Z^{\ju} \cdot Z^{\ku} :=Z_I{}^{\ju} Z_I{}^{\ku} =0~, \qquad 
\forall {\ju}, {\ku} = 1, \dots, m~;
\eea
and (iii) be defined modulo the equivalence relation 
\bea
Z_I{}^{\ju} ~\sim ~Z_I{}^{\ku} \, D_{\ku}{}^{\ju} ~, \qquad D =(D_{\ku}{}^{\ju} ) \in \sGL (m, {\mathbb C})~.
\label{6.5}
\eea
In the case $\cN=3$ and $m=1$,  ${\mathbb M}{}^{3|6} \times {\mathbb X}^3_1$
may be seen to be  equivalent to the standard $\cN=3$ harmonic superspace 
${\mathbb M}{}^{3|6} \times {\mathbb C}P^1$
\cite{ZH}. It was shown in  \cite{KPT-MvU} 
that for $\cN>2$ and $m=1$ the internal manifold  ${\mathbb X}^\cN_1$ is a symmetric 
space, 
\bea
{\mathbb X}^\cN_1= \sSO (\cN) / \sSO (\cN-2) \times \sSO(2) ~, \qquad \cN>2~.
\eea

When dealing with the $\cN=5$ supercurrent $J^I$, it is natural to 
make use of the harmonic/projective superspace 
 ${\mathbb M}{}^{3|10} \times {\mathbb X}^5_1$. 
 Using the null five-vector $Z_I$ parametrising 
$  {\mathbb X}^5_1$, we introduce the first-order operators
\bea
{\mathfrak D}_\alpha := Z_I D^I_\alpha~, 
\qquad \{ {\mathfrak D}_\alpha , {\mathfrak D}_\b \} =0
\eea
and associate with $J^I$ the superfield  ${\mathfrak J}:= Z_I J^I$.  
Then, the supercurrent conservation equation \eqref{J-eq}
implies that  ${\mathfrak J}$ is an analytic superfield, 
\bea
{\mathfrak D}_\alpha {\mathfrak J} =0~.
\eea

When dealing with the $\cN=6$ supercurrent $J^{IJ} = - J^{JI}$, it is natural to 
make use of the harmonic/projective superspace 
 ${\mathbb M}{}^{3|12} \times {\mathbb X}^6_2$. 
 Using the null six-vectors $Z_I{}^{\ju}$ parametrising $ {\mathbb X}^6_2$, 
 we introduce the first-order operators
 \bea
 {\mathfrak D}^{\ju}_{\alpha}:= Z_I{}^{\ju} D^I_\a~, 
\qquad \{ {\mathfrak D}^{\ju}_\alpha , {\mathfrak D}^{\ku}_\b \} =0~, \qquad
\ju, \ku = \1, \2
\eea
and associate with $J^{IJ}$ the superfield 
${\mathfrak J}  := \hf \ve_{\ju \ku} Z^{\ju}_I Z^{\ku}_J J^{IJ}$, 
with $\ve_{\ju \ku} $ an antisymmetric tensor. 
Then, the supercurrent conservation equation (\ref{N6-conserv}) 
implies that ${\mathfrak J} $ is an analytic superfield, 
\bea
{\mathfrak D}^{\ju}_{\alpha} {\mathfrak J} =0~.
\eea

It is clear that the correlation functions of the $\cN=5$ and $\cN=6$
supercurrent multiplets should simplify if the above harmonic/projective superspace settings are used. It would be interesting to develop superconformal 
formalisms to compute correlation functions of primary analytic superfields 
in such superspaces. 

In a recent work \cite{LMM}, a non-standard $\cN=6$ harmonic superspace was introduced with the aim to study the three-point correlation functions of BPS operators in $\cN=6$ superconformal field theories. It was pointed out that the supercurrent multiplet 
is a BPS operator, and therefore \cite{LMM} provided a harmonic-superspace  expression for the supercurrent three-point correlator. However, 
the authors of  \cite{LMM} did not describe how their harmonic superspace is related 
 to the superspaces ${\mathbb M}{}^{3|12} \times {\mathbb X}^6_m$
introduced  in  \cite{KPT-MvU}. 
As a result, a precise relationship between the results of \cite{LMM} 
and the present paper remains to be understood. 

Another possible extension of the present work is the study of four-point correlation functions of conserved currents in three-dimensional superconformal field theories. One can hope that the extended supersymmetry imposes so strong constraints, 
for sufficiently large $\cN$,  on the four-point correlators that their form can be found explicitly, similarly as it was demonstrated for the 4D $\cN=4$ SYM theory \cite{KS}. We leave these issues for further studies.
\\

\noindent
{\bf Acknowledgements:}\\
We are grateful to Evgeny Buchbinder for useful discussions. This work is supported in part by the ARC DP project DP140103925. 

\appendix

\section{Two- and three-point building blocks}

In this appendix we give a brief summary of the two- and three-point 
superconformal structures $\cN$-extended superspace, which were introduced 
in \cite{BKS1}. These structures have been used in the construction of 
the supercurrent correlation functions in the main body on the paper. 

Consider $\cN$-extended Minkowski superspace
${\mathbb M}{}^{3|2\cN}$
parametrised by real bosonic $x^{\a\b}=x^{\b\a}$ and fermionic $\q^\a_I$
coordinates
\be
z^A=(x^{\alpha\beta},\theta^{\alpha}_I),  \qquad
\a = 1,2~, \qquad I = 1, \dots, \cN~.
\label{ss-coord}
\ee
Here  $\a,\b$ are the $\sSL(2,\mathbb{R})$ spinor indices, while $I$ is the $R$-symmetry index.
All building blocks are composed of the
two-point structures
\begin{subequations}
\bea
{\bm x}_{12}^{\alpha\beta} &=& (x_{1}-x_{2})^{\alpha\beta}
+2\ri \theta_{1I}^{(\alpha}\theta_{2I}^{\beta)}
-\ri\theta_{12I}^\alpha \theta_{12I}^\beta~,\label{super-interv-X}\\
\theta_{12I}^\alpha &=& (\theta_{1}-\theta_2)^\alpha_{I}~.
\label{super-interv-Theta}
\eea
\label{super-interv}
\end{subequations}
The matrix \eqref{super-interv-X} has the following
symmetry property
\be
{\bm x}_{21}^{\alpha\beta} = -{\bm x}_{12}^{\beta\alpha}~.
\label{antisymmetry}
\ee
A useful object is the square of this matrix
\be
{\bm x}_{12}{}^2 := -\frac12{\bm x}_{12}^{\alpha\beta}{\bm
x}_{12\alpha\beta}~.
\label{4.133}
\ee

One more important two-point structure is the 
$\cN\times \cN$ matrix
\be
u_{12} =(u_{12}^{IJ})~, \qquad
u_{12}^{IJ} = \delta^{IJ} + 2\ri
\theta^{ \a I}_{12}({\bm x}_{12}^{-1})_{\alpha\beta}
\theta^{\b J}_{12}~,
\label{two-point-u}
\ee
where
\be
({\bm x}_{12}^{-1})_{\alpha\beta} = -\frac{
{\bm x}_{12\beta\alpha}}{{\bm x}_{12}{}^2}
\ee
is the inverse for $({\bm x}_{12})^{\alpha\beta}$,
that is $({\bm x}_{12}^{-1})_{\alpha\beta}({\bm x}_{12})^{\beta\gamma}
=\delta_\alpha^\gamma$.
One may check that the matrix
$u_{12}$ is orthogonal and unimodular,
\be
u_{12}^{\rm T}u_{12} = {\mathbbm 1}_\cN~,\qquad
\det u_{12} = 1~.
\label{2.17}
\ee
As is shown in \cite{BKS1}, the two-point structures (\ref{super-interv-X}),  (\ref{4.133}) and (\ref{two-point-u}) transform covariantly under the superconformal group i.e., as the tensors with Lorentz an $\sSO(\cN)$ indices at both superspace points. Here we do not give their transformation laws referring the readers to our previous works \cite{BKS1,BKS2}.

Associated with three superspace points $z_1$, $z_2$ and $z_3$
are the following three-point structures:
\begin{subequations}
\label{three-points}
\bea
{\bm X}_{1\alpha\beta}&=&-({\bm x}^{-1}_{21})_{\alpha\gamma}
{\bm x}_{23}^{\gamma\delta}
({\bm x}^{-1}_{13})_{\delta\beta}~,\label{3ptX}\\
\Theta^I_{1\alpha}&=& ({\bm x}^{-1}_{21})_{\alpha\beta}\theta^{I\beta}_{12}
-({\bm x}^{-1}_{31})_{\alpha\beta}\theta_{13}^{I\beta}~,\label{3ptTheta}\\
U_{1}^{IJ}&=&u_{12}^{IK}u_{23}^{KL}u_{31}^{LJ}~.
\label{U}
\eea
\end{subequations}
These objects are labeled by the index $1$ reflecting the fact that they transform as tensors in the superspace point $z_1$. Performing the cyclic permutation of superspace points $z_1$, $z_2$ and $z_3$ one can obtain similar objects transforming as tensors in points $z_2$ and $z_3$. The three-point structures at different superspace
points are related to each other as follows
\begin{subequations}
\label{222}
\bea
{\bm x}_{13}^{\alpha\alpha'}{\bm X}_{3\alpha'\beta'}{\bm x}_{31}^{\beta'\beta}
&=&-({\bm X}_{1}^{-1})^{\beta\alpha}=\frac{{\bm X}_{1}^{\alpha\beta}}{{\bm
X}_{1}{}^2}~,\label{222a}\\
\Theta^I_{1\gamma} {\bm x}_{13}^{\gamma\delta}{\bm X}_{3\delta\beta}
&=&u^{IJ}_{13} \Theta^J_{3\beta}~,\label{222b}\\
U_{3}^{IJ} &=& u_{31}^{IK}U_{1}^{KL}u_{13}^{LJ}~.
\label{222c}
\eea
\end{subequations} 

The three-point structures (\ref{three-points}) have several important properties. In particular, the tensor
(\ref{3ptX}) can be decomposed into symmetric and antisymmetric
parts
\be
{\bm X}_{\alpha\beta} = X_{\alpha\beta}
-\frac{\ri}2\varepsilon_{\alpha\beta}\Theta_{}^2~,
\label{XX}
\ee
where  the symmetric spinor
$X_{\alpha\beta}=X_{\beta \a}$ is equivalently represented as
a three-vector $X_{m}=-\frac12 \gamma_m^{\alpha\beta}X_{\alpha\beta}$. Here and below we suppress the subscript labelling the superspace point.

Next, the matrix (\ref{U}) can be expressed in terms of
(\ref{3ptX}) and (\ref{3ptTheta}) similarly to
(\ref{two-point-u}):
\be
U^{IJ}=\delta^{IJ}
+2\ri \Theta^I_{\alpha}({\bm X}^{-1})^{\alpha\beta}\Theta^J_{\beta}
=\delta^{IJ}-2\frac{A^{IJ}}{{\bm X}^2}
+\frac{\Theta^{I\alpha}\Theta^J_\alpha \Theta^2}{{\bm X^2}}
~,
\label{U-explicit}
\ee
where
\be
A^{IJ} = \ri \Theta^{I\alpha}X_{\alpha\beta} \Theta^{J\beta}~.
\label{A}
\ee

We point out that the three-point objects (\ref{three-points}) look like local expressions. In fact, in computing correlators we  consider functions of these objects obeying certain differential equations. These differential equations involve generalised superspace derivatives such as
\be
{\cal D}^I_{\alpha} = \frac\partial{\partial \Theta_{I}^\alpha }
 +\ri\gamma^m_{\alpha\beta}\Theta^{I\beta} \frac\partial{\partial X^m}~.
\label{generalized-DQ}
\ee
This derivative should not be confused with the usual superspace derivative $D^I_\alpha = \frac\partial{\partial \theta_{I}^\alpha }
 +\ri\gamma^m_{\alpha\beta}\theta^{I\beta} \frac\partial{\partial x^m}$ which acts on the superspace coordinates (\ref{ss-coord}).
The anticommutation relations for these derivatives are
\be
\{ D^I_\alpha , D^J_\beta \} = 2\ri \delta^{IJ}\partial_{\alpha\beta}~,
\label{D-comm}
\ee 
and similar for the generalised ones (\ref{generalized-DQ}).

\begin{footnotesize}

\end{footnotesize}

\end{document}